\documentclass[12pt]{article}
\usepackage{color,rotate,graphics,epsf,psfrag,amssymb}
\usepackage{hyperref}

\newcommand{\insertfig}[2]{\mbox{\epsfxsize=#1cm \epsfbox{#2.eps}}}

\newcommand{\GeV}{{\rm GeV}}
\font\cmss=cmss12 
\def\1{\hbox{{1}\kern-.25em\hbox{l}}}
\def\bfZ{\relax{\hbox{\cmss Z\kern-.4em Z}}}

\begin{document}

\begin{titlepage}

\centerline{\large \bf Next--to--next--to--leading order corrections
to} \vspace{2mm} \centerline{\large \bf deeply virtual Compton
scattering: the non--singlet case}

\vspace{15mm}

\centerline{\bf  D.~M\"uller$$}

\vspace{10mm}

\centerline{\it Institut f\"ur Theoretische Physik, Universit\"at
Regensburg} \centerline{\it D--93040 Regensburg, Germany}

\vspace{10mm}

\centerline{\it Department of Physics and Astronomy, Arizona State University}
\centerline{\it P.O. Box 871504, Tempe, AZ 85287--1504, USA}

\vspace{10mm}

\centerline{\bf Abstract}

\vspace{0.5cm}

\noindent We study perturbative QCD corrections to deeply virtual
Compton scattering on an unpolarized nucleon target in the flavor
non--singlet sector to next--to--next--to--leading order accuracy,
restricting ourselves to the kinematically dominant amplitude.
The difference between the standard minimal subtraction and the
conformal scheme, in which conformal symmetry is manifest, is
studied to next--to--leading order. Beyond this order we employ
conformal symmetry for the evaluation of perturbative corrections.
Within a certain class of generalized parton distributions we find
moderate radiative  corrections.\\

\vspace{50mm}

\noindent Keywords: perturbative corrections, deeply virtual
Compton scattering, conformal symmetry

\vspace{0.5cm}

\noindent PACS numbers: 11.10.Hi, 12.38.Bx, 13.60.Fz

\end{titlepage}

\section{Introduction}

The hard--exclusive photo electroproduction off a nucleon, among a
class of hard exclusive but inelastic processes, is generally
considered as the theoretic cleanest process to gain access to the
so--called generalized parton distributions (GPDs)
\cite{MueRobGeyDitHor94,Ji96}. These non--perturbative
distributions contain manifold information that cannot be gained
from measurements of deep (semi--)inelastic or elastic and other
exclusive processes \cite{BroLep80}. Based on a partonic
interpretation, GPDs allow us to reveal the internal structure of
the probed hadron, especially of the nucleon, from a new
perspective, for comprehensive reviews see Ref.\
\cite{Die03aBelRad05}.

We recall that the process in question has two interfering
subprocesses, namely, the Bethe--Heitler brems\-strahlung and
deeply virtual Compton scattering (DVCS). The former is known in
terms of the electromagnetic form factors and so the latter can be
measured in several asymmetries, see e.g., Ref.\ \cite{DVCS}, that
appear in leading order of the expansion with respect to the
inverse photon virtuality, i.e., single and double spin and charge
asymmetries
\cite{DieGouPirRal97,BelMueNieSch00,BelMueKir01,KirMue03}. Relying
on the validity of the operator product expansion (OPE) of two
electromagnetic currents, the DVCS amplitude factorizes to leading
power, i.e., twist--two, in short-- and long--distance physics, where
the former and latter are incorporated in (resummed)
Wilson--coefficients and GPDs, respectively
\cite{MueRobGeyDitHor94}. Equivalently, in the partonic framework,
the DVCS amplitude to leading power accuracy is represented as
convolution of hard--scattering amplitudes, systematically
calculable in powers of the strong coupling, and GPDs. The
factorization of collinear singularities was shown to all orders
in perturbation theory \cite{ColFre98}.

Let us remind that for experimental accessible scales, the
perturbative description of elastic exclusive processes is
controversially debated \cite{IsgLle89a}. With respect to this
fact one might wonder whether the perturbative GPD framework is
justified. For the GPD phenomenology it is essential to clarify
this issue; unfortunately, this is an interlaced task.  The
relative size of perturbative and non--perturbative, i.e., power
suppressed contributions,  might serve as criteria for the
reliability of the perturbative framework. The DVCS
hard--scattering amplitudes (Wilson--coefficients) were
perturbatively evaluated up to next--to--leading order (NLO) in the
strong coupling \cite{BelMue97a,ManPilSteVanWei98}. It turned out
that these radiative corrections  can be of the order of 30\% to
50\% for fixed target kinematics \cite{BelMueNieSch99}. Even much
larger corrections have been reported for the kinematics of
collider experiments \cite{FreMcD01b}. Mainly, but not only, this
is due the appearances of gluonic GPDs at NLO, see also discussion
in Sect.\ 6.2.3 of Ref.\ \cite{BelMueKir01}. The problem arises
whether these sizeable corrections are related to the appearance
of gluons, are induced by an awkward choice of GPD ans\"atze, or
indicate that at experimental accessible scales the perturbative
regime has not be fully set in.

To get a deeper insight into this issue,  we study in this letter
perturbative corrections to the DVCS process beyond NLO accuracy
within a framework that avoids a cumbersome diagrammatic
calculation. Based on the conformally covariant OPE (COPE),  the
perturbative corrections up to  next--to--next--to--leading order
(NNLO) can be economically evaluated  \cite{Mue97a,BelMue97a}, see
also  review \cite{BraKorMue03}. Thereby we restrict ourselves to
the flavor non--singlet part of the helicity conserved twist--two
amplitude in the parity even sector, which is the dominant
contribution in several DVCS observables. In Sect.\
\ref{Sec-GenFor} we outline the  evaluation of this amplitude by
means of the COPE \cite{Mue97a,BelMue97a,MelMuePas02}. This yields
a divergent series of conformal GPD moments, which is resummed
within a Mellin--Barnes integral. In Sect.\ \ref{Sec-NumRes} we
numerically study the radiative corrections, mainly due to the
Wilson--coefficients, at NLO and NNLO. Especially, we explore the
numerical differences of the NLO corrections in the modified
minimal subtraction ($\overline{\rm MS}$) and conformal
subtraction (CS) schemes. We then provide  estimates to NNLO
accuracy. Finally, in Sect.\ \ref{Sec-Con} we conclude.

\section{General formalism}
\label{Sec-GenFor}

Let us first recall the standard perturbative QCD framework.
Usually, one employs the $\overline{\rm MS}$ scheme, which is
based on dimensional regularization and the removal of the poles
with respect to the dimensional parameter $(4-n)/2$. This scheme
is used twofold: (i) to define the renormalized strong coupling
and (ii) for the factorization of collinear singularities or if
one wishes  for the renormalization of (leading twist) composite
operators\footnote{To leading twist approximation there is an
one--to--one cross talk between the ultra--violet and collinear
singularities in the partonic matrix elements of these
operators.}, which are labelled by the good quantum number spin.
In exclusive reactions such operators with given spin are also be
build within total derivatives. In general operators with the same
spin will mix under renormalization. To get rid of this mixing
phenomenon, one changes at some stage of the full calculation in an
explicit or implicit manner to a basis of multiplicatively
renormalizable operators, i.e., one diagonalizes the evolution
equation \cite{EfrRad80}. For vanishing $\beta(\alpha_s)$ function
this  transformation leads to conformal
operators \cite{Mue94}.  Such operators  transform covariantly
under the so--called collinear conformal transformation
$SL(2,\mathbb R)$ and are members of infinite dimensional
conformal multiplets (towers) that are labelled by the conformal
spin. With other words one takes explicitly or implicitly
advantages from the underlying conformal symmetry of the classical
QCD Lagrangian. If the trace anomaly of the energy--momentum
tensor, proportional to $\beta(\alpha_s)$, is absent, conformal
symmetry is present in perturbative QCD and its predictive power
can be employed at any order.

In the following we evaluate the scattering amplitude for the DVCS
process, which is expressed in terms of the time ordered product
of two electromagnetic currents
\begin{eqnarray}
\label{Def-HadTen} T_{\mu\nu} (q, P_1, P_2) = \frac{i}{e^2} \int
dx e^{i x\cdot q} \langle P_2, S_2 | T j_\mu (x/2) j_\nu (-x/2) |
P_1, S_1 \rangle,
\end{eqnarray}
where $q = (q_1 + q_2)/2$ ($\mu$  and $q_2$ refers to the outgoing
real photon). The incoming photon has a large virtuality
$q_1^2=-{\cal Q}^2$ and requiring that in the limit $-q^2 = Q^2\to
\infty$ the scaling variables
\begin{eqnarray}
\xi = \frac{Q^2}{P\cdot q}\,,\qquad \eta = -\frac{\Delta\cdot
q}{P\cdot q}\,,\qquad P = P_1 + P_2\,, \qquad \Delta=P_2-P_1\,.
\end{eqnarray}
and the momentum transfer squared $\Delta^2$  are fixed, the
dominant contributions arise from the light--cone singularities of
the time ordered product. In this generalized Bjorken limit one
can now employ the OPE to evaluate the hadronic tensor in terms of
the leading twist--two operators, where for DVCS kinematics
$\eta\simeq \xi$ and $Q^2 \simeq {\cal Q}^2/2$.

Before we outline this step let us introduce a parameterization of
the hadronic tensor
\begin{eqnarray}
\label{decom-T} T_{\mu\nu} (q,P,\Delta) &=& - \tilde{g}_{\mu\nu}
\frac{q_\sigma V^\sigma}{P\cdot q} - i \tilde{\epsilon}_{\mu \nu \rho
\sigma} \frac{q^\rho A^\sigma}{P\cdot q} + \cdots.
\end{eqnarray}
To ensure current conservation, the metric and Levi--Civita tensors
are contracted here with projection operators, for explicit
definitions of $\tilde g_{\mu\nu}$ and $\tilde
\epsilon_{\mu\nu\rho\sigma}$ see, e.g., Ref.\
\cite{BelMueNieSch00}. The ellipsis indicates terms that are
finally power suppressed in the DVCS amplitude or a determined by
the gluon transversity GPD, which is not considered here. We note
that in the forward limit $\Delta\to 0$ the first and second term
on the r.h.s.\ are expressed by  the deep inelastic structure
functions $F_1$ and $g_1$. In the parity even sector the vector
\begin{eqnarray}
\label{dec-FF-V} V^{\sigma} = \bar U (P_2, S_2) \left( {\cal H}
\gamma^\sigma + {\cal E} \frac{i\sigma^{\sigma\rho}
\Delta_\rho}{2M} \right) U (P_1, S_1) + \cdots\, ,
\end{eqnarray}
is decomposed in Compton form factors (CFFs) ${\cal H}$ and ${\cal
E}$, similar for the axial--vector $A^{\sigma}$ in terms of
$\widetilde{\cal H}$ and $\widetilde{\cal E}$, where again higher
twist contributions are neglected.

Now we are in the position to employ the OPE.  Let us first
suppose that the trace anomaly is absent and so conformal symmetry
can be employed. Formally, this can be achieved by assuming that a
hypothetical fixed point exist, i.e., $\beta(\alpha_s^\ast)=0$. We
can then use the COPE, which tells us how the total derivatives
are arranged. Moreover, it can be considered as a partial wave
expansion with respect to the conformal spin $j+2$. For the
time--ordered product of two electromagnetic currents it reads in
the flavor non--singlet sector as \cite{FerGriGat71,Mue97a}
\begin{eqnarray}
\label{Def-COPE}
 \left\{T j_\mu (x) j_\nu (0)\right\}^{\rm NS} &\!\!\! =\!\!\! &
\frac{1}{\pi^2} \frac{-\tilde g_{\mu \nu}}{(-x^2 + i 0)^2}
\sum_{j=1}^\infty \left[1- (-1)^j\right]
\frac{\Gamma(2-\gamma_j/2)(2+j+\gamma_j/2)}{
2^{\gamma_j/2}\Gamma(j+1+\gamma_j/2)}
 C_j^{\rm NS}([-x^2 + i 0] \mu^2)
\nonumber\\
&&\hspace{3cm}\times (-2 i x_-)^{j+1} \int_{0}^{1}\!du\,
[u(1-u)]^{j+1+\gamma_j/2} {\cal O}^{\rm NS}_j(u x_-) + \cdots\,,
\end{eqnarray}
where $x_- = \widetilde n \cdot x $ with $\widetilde n^2=0$ is the
projection of the vector $x$ on the light cone.  The anomalous
dimensions of the multiplicatively renormalizable operators ${\cal
O}^{\rm NS}_j$ are denoted as $\gamma_j(\alpha_s^\ast)$. The advantage
of the COPE is that the Wilson--coefficients (electrical charge
factors will be omitted)
\begin{eqnarray}
\label{Def-WilCoeC} C_j^{\rm NS}(-x^2 \mu^2) =
\left(-\mu^2 x^2\right)^{\gamma_j/2}
\frac{2^{j+1+\gamma_j/2}\Gamma(5/2+j+\gamma_j/2)}{
\Gamma(3/2)\Gamma(3+j+\gamma_j/2)}\,
c_j(\alpha_s^\ast)\,,
\end{eqnarray}
are up to the normalization $c_j(\alpha_s^\ast)$ known. The conformal
operators
\begin{eqnarray}
\label{Def-ConOpe} {\cal O}^{\rm NS}_j(u x_-) =
\frac{\Gamma(3/2)\Gamma(1+j)}{2^{j} \Gamma(3/2+j)}\,  (i
\partial_+)^j \bar{\psi}(u x_-) { \gamma_+}\, C_j^{3/2}\!\left(\!
\frac{\!\stackrel{\leftrightarrow}{D}_{+}}{\partial_+}\!
\right)\psi(u x_-)\,,
\end{eqnarray}
are defined with a non--standard normalization. Here
$\stackrel{\leftrightarrow}{D}_+ = \stackrel{\rightarrow}{D}_+
-\stackrel{\leftarrow}{D}_+$ and $\partial_+ =
\stackrel{\rightarrow}{D}_+ + \stackrel{\leftarrow}{D}_+$ are the
covariant and total derivatives, contracted with the light-like
vector $n$, i.e., $n^2=0$ and $n\cdot\widetilde n =1$, $C_j^{3/2}$
are the Gegenbauer polynomials of order $j$ with index $3/2$. The
normalization is chosen so that in the forward limit $\Delta \to
0$ the reduced matrix elements of the conformal operators
\begin{eqnarray}
\label{Def-ConMomVec} 
\frac{1}{P_+^{j+1}} \langle P_2, S_2 \big|  {\cal O}_j(0) \big|P_1,
S_1 \rangle= \frac{1}{P_+}\bar U (P_2, S_2)\! \left(\! H_j(\eta,\Delta^2,\mu^2)
\gamma_+ + E_j(\eta,\Delta^2,\mu^2) \frac{i\sigma_{+\nu}
\Delta^\nu}{2M}\! \right) \! U (P_1, S_1)
\end{eqnarray}
coincide with the Mellin moments of parton densities. Especially,
for odd values of $j$ the forward limit of $H_j$ is the sum of the
quark and anti--quark  density moments, i.e., $\lim_{\Delta\to
0}H_j=q_j+\overline{q}_j$. Moreover, in this limit the
$u$--integration in Eq.\ (\ref{Def-COPE}) can be performed and
leads to the well--known OPE that is used for the deep inelastic
scattering structure function $F_1$.  Hence, $c_j(\alpha_s^\ast)$
are identified as the Wilson--coefficients of $F_1$, known to NNLO
order.

Employing the COPE (\ref{Def-COPE}), we can now achieve the
factorization of the CFFs (\ref{dec-FF-V}) in a straightforward
manner. Plugging Eq.\ (\ref{dec-FF-V}) into the definition of the
hadronic tensor (\ref{Def-HadTen}), performing Fourier transform,
and form factor decomposition (\ref{dec-FF-V}), we arrive for the
DVCS kinematics $\eta=\xi$ at \cite{Mue97a}
\begin{eqnarray}
\label{Def-ConParDecInt} \left\{{{\cal H}^{\rm NS} \atop {\cal
E}^{\rm NS}} \right\}
= \sum_{j=1}^\infty \frac{1-(-1)^j}{2} {\xi}^{-j-1} C^{\rm
NS}_j(\mu^2/Q^2) \left\{{H_j^{\rm NS} \atop E_j^{\rm NS}} \right\}
(\eta, \Delta^2,\mu^2)\Big|_{\eta=\xi}\,.
\end{eqnarray}
Here $H_j^{\rm NS}$ and $E_j^{\rm NS}$, cf.\ Eq.\ (\ref{Def-ConMomVec}),
are given by the conformal moments of the corresponding GPDs
\begin{eqnarray}
\label{Def-ConGPDmom}
\left\{{H_j^{\rm NS} \atop E_j^{\rm NS}} \right\} (\eta,
\Delta^2,\mu^2)= \frac{\Gamma(3/2)\Gamma(1+j)}{2^{j}
\Gamma(3/2+j)} \int_{-1}^{1 }\! dx\,\eta^j
C_j^{3/2}\!\left(\frac{x}{\eta}\right) \left\{{H^{\rm NS} \atop
E^{\rm NS}} \right\} (x,\eta, \Delta^2,\mu^2)\,.
\end{eqnarray}
Note that these expectation values are
measurable on the lattice \cite{Hagetal03}.

Unfortunately, the series (\ref{Def-ConParDecInt}) is divergent
for $|\xi| < 1$ and must be resummed, e.g., by means of the
Sommerfeld--Watson transformation \cite{MueSch05}. Alternatively,
using the trick that $\eta$ is not equated to $\xi$ allows to deal
with Eq.\ (\ref{Def-ConParDecInt}) in the analogous manner  as it
is known from deep inelastic scattering. Namely, a dispersion
relation allows to express the Mellin moments of the imaginary
part by the partial waves, appearing in the COPE
(\ref{Def-ConParDecInt}), and thus the inverse  Mellin transform
provides the imaginary part of the CFFs. The real
part can be restored from the dispersion relation, too, and
finally equating $\eta=\xi$ leads to the same representation as
presented in Ref.\ \cite{MueSch05}:
\begin{eqnarray}
\label{Def-MelBar} \left\{{{\cal H}^{\rm NS} \atop {\cal E}^{\rm
NS}} \right\}(\xi,\Delta^2,\mu^2)
=  \frac{1}{2 i}\int_{c-i \infty}^{c+ i \infty}\! dj\, \xi^{-j-1}
\left[i +\tan\left(\frac{\pi j}{2}\right) \right] C_j^{\rm
NS}(Q/\mu) \left\{{H_j^{\rm NS} \atop E_j^{\rm NS}} \right\} (\xi,
\Delta^2,\mu^2)\,,
\end{eqnarray}
where all singularities of the conformal GPD moments and conformal
moments lie on the l.h.s.\ of the integration path $(c<1)$. Note,
however, that the analytic continuation of  the
Wilson--coefficients (\ref{Def-WilCoeC}) with respect to the
conformal spin, analogously done as in deep inelastic scattering,
leads essentially to an exponential  $2^j$ growing with increasing
$j$. For $\xi> 1$, this must be weighed down by a suppression
factor that comes from the conformal moments
(\ref{Def-ConGPDmom}). This is a rather nontrivial requirement on
their analytic continuation. It should be done in such a way that
the integral (\ref{Def-MelBar}) remains unchanged if the
integration contour is closed by an infinite arc, surrounding  the
first and forth quadrant.  The residue theorem states then that
this Mellin--Barnes integral is for $\xi> 1$ equivalent to the
series (\ref{Def-ConParDecInt}). We note that the form of the
integrand (\ref{Def-MelBar}) does in fact not rely on conformal
symmetry and so it can be used in any scheme within the
corresponding Wilson--coefficients.

In the $\overline{\rm MS}$ scheme the conformal symmetry is not
manifestly implemented in the OPE, however, this failure can be
cured by a finite renormalization \cite{Mue97a,BelMue97a}. As
explained above, in such a CS scheme and for $\beta=0$ we can
simply borrow the Wilson--coefficients $c_j(\alpha_s)$ and
anomalous dimensions $\gamma_j(\alpha_s)$ from deep inelastic
scattering. The inclusion of $\beta$ proportional terms in Eq.\
(\ref{Def-MelBar}) is, as factorization by itself, conventional.
Two possibilities have been discussed in Ref.\ \cite{MelMuePas02}.
Namely, in the CS scheme, one might add the $\beta$ proportional
term from the $\overline{\rm MS}$ scheme, evaluated to NNLO in
Ref.\ \cite{BelSch98}, while in the $\overline{\rm CS}$ scheme the
running of the coupling is implemented in the form of the COPE
(\ref{Def-COPE}) in such a way that the Wilson--coefficients
autonomously evolve in the considered order \cite{MelMuePas02}. In
this letter we prefer the latter convention and, moreover, will
equate the factorization and renormalization scales, i.e,
$\mu=\mu_f=\mu_r$. After expansion with respect to $\alpha_s$, we
write, discarding in the following the superscript NS,  the
Wilson--coefficients (\ref{Def-WilCoeC}) as
\begin{eqnarray}
\label{Res-WilCoe-Exp-CS}
C_j
&\!\!\! =\!\!\!  & \frac{2^{j+1}
\Gamma(j+5/2)}{\Gamma(3/2)\Gamma(j+3)} \left[{\cal C}_j^{(0)} +
\frac{\alpha_s(\mu)}{2\pi} {\cal C}_j^{
(1)}({\cal Q}/\mu)
+ \frac{\alpha^2_s(\mu)}{(2\pi)^2} {\cal C}_j^{
(2)}({\cal Q}/\mu) + {\cal O}(\alpha_s^3) \right]\,,
\\
\label{Res-WilCoe-CS-NLO}
{\cal C}_j^{
(0)} &\!\!\! =\!\!\! & 1\,,\qquad
 {\cal C}_j^{
(1)} 
= c_j^{
(1) }+
\frac{\gamma_j^{
(0)}}{2} \left[s^{(1)}_j+\ln\frac{\mu^2}{{\cal Q}^2}\right]\,,
\\
\label{Res-WilCoe-CS-NNLO}
{\cal C}_j^{
(2)} &\!\!\! =\!\!\!  & c_j^{
(2)}+
\frac{\gamma_j^{
(1)}}{2} \left[ s^{(1)}_j +
\ln\frac{\mu^2}{{\cal Q}^2}\right] + \frac{\gamma_j^{
(0)}}{2} \left[ c_j^{
(1)} \left\{s^{(1)}_j +
\ln\frac{\mu^2}{{\cal Q}^2}\right\} + \frac{\gamma_j^{
(0)}}{4} \right.
\\
&&
 \times\left.\left\{s^{(2)}_j + 2 s^{(1)}_j \ln\frac{\mu^2}{{\cal
Q}^2}+ \ln^2\frac{\mu^2}{{\cal Q}^2} \right\}\right]
-\frac{\beta_0}{2}
 \left[ c_j^{
(1)}+
\frac{\gamma_j^{
(0)}}{2} s_j^{(1)} + \frac{\gamma_j^{
(0)}}{4} \ln\frac{\mu^2}{{\cal Q}^2} \right]\ln\frac{\mu^2}{{\cal
Q}^2}\,,
 \nonumber
\end{eqnarray}
where
\begin{eqnarray}
\gamma_j^{
(0)}&\!\!\!=\!\!\!& C_F \left(4S_1(j+1)- 3
-\frac{2}{(j+1)(j+2)}  \right)\,, \quad C_F= \frac{4}{3}\,,\
\\
c_j^{
(1)}&\!\!\!=\!\!\!& C_F \left[S^2_{1}(1 + j) +
\frac{3}{2} S_{1}(j + 2) - \frac{9}{2}  + \frac{5-2S_{1}(j)}{2(j +
1)(j + 2)} -   S_{2}(j + 1)\right]\,,
\\
s_j^{(1)}&\!\!\!=\!\!\!& S_1(j+3/2)-S_1(j+2) + 2 \ln(2)\,, \quad
s_j^{(2)}= \left(s_j^{(1)}\right)^2 -S_2(j+3/2)+ S_2(j+2) \,.
\end{eqnarray}
The analytic continuation of the harmonic sums are defined by
$S_1(z)= d\ln \Gamma(z+1)/dz +\gamma_E$ and $S_2(z)= -d^2\ln
\Gamma(z+1)/dz^2 + \zeta(2)$, where $\gamma_E$ is the Euler
constant. The first expansion coefficient of $\beta(g)/g =
(\alpha_s/4\pi) \beta_0 + {\cal O}(\alpha_s^2)$ is $\beta_0 =
(2/3) n_f-11$, where $n_f$ is the number of active quarks. The
two--loop quantities $c_j^{(2)}$ and $\gamma_j^{(1)}$ are lengthy
and can be obtained from Ref.\ \cite{CurFurPet80,ZijNee92}. The
evolution of the flavor non--singlet (integer) conformal moments in
this $\overline{\rm CS}$ scheme is  governed by
\begin{eqnarray}
\label{Def-RGE} \mu\frac{d}{d\mu} \left\{{ H_{j}^{\rm NS}\atop
E_{j}^{\rm NS}}\right\}(\eta, \Delta^2,\mu^2) &\!\!\!=\!\!\!& -\Bigg[
\frac{\alpha_s(\mu)}{2\pi} \gamma_j^{(0)} +
\frac{\alpha_s^2(\mu)}{(2\pi)^2} \gamma_j^{(1)}+
\frac{\alpha_s^3(\mu)}{(2\pi)^3} \gamma_j^{(2)} +{\cal
O}(\alpha_s^4) \Bigg] \left\{{ H_{j}^{\rm NS}\atop E_{j}^{\rm NS}}\right\}(\eta,
\Delta^2,\mu^2)
\nonumber\\
&&\hspace{0.5cm} -\frac{\beta_0}{2}
\frac{\alpha_s^3(\mu)}{(2\pi)^3}\sum_{k=0}^{j-2} \eta^{j-k}
\left[\Delta_{jk}^{\overline{{\rm CS}}}+{\cal O}(\alpha_s)
\right]\left\{{ H_{k}^{\rm NS}\atop E_{k}^{\rm NS}}\right\}(\eta, \Delta^2,\mu^2)\,,
\end{eqnarray}
where the mixing matrix $\Delta_{jk}^{\overline{{\rm CS}}}$ is not
completely known.

We remark that the NLO corrections in the $\overline{\rm MS}$
scheme can be easily evaluated from the conformal moments of the
hard--scattering amplitude, e.g., given in Ref.\
\cite{BelMueNieSch99}. All integrals, which are needed, are given
in Appendix C of Ref.\ \cite{MelMuePas02} for integer conformal
spin. The analytic continuation is straightforward and so in the
$\overline{\rm MS}$ scheme Eq.\ (\ref{Res-WilCoe-CS-NLO}) is to
replace by
\begin{eqnarray}
\label{Res-WilCoe-MS-NLO} {\cal C}_j^{ {\overline{\rm MS}}
(1)}&\!\!\!=\!\!\!& C_F \left[2 S^2_{1}(1 + j)- \frac{9}{2}  +
\frac{5-4S_{1}(j+1)}{2(j + 1)(j + 2)} +
\frac{1}{(j+1)^2(j+2)^2}\right] + \frac{\gamma_j^{
(0)}}{2}
\ln\frac{\mu^2}{{\cal Q}^2} \,.
\end{eqnarray}
In this scheme also the complete anomalous dimension matrix is
known to two--loop accuracy \cite{Mue94}.

\section{Numerical results}
\label{Sec-NumRes}

In this Section we numerically study the   radiative corrections
to the CFF
\begin{eqnarray}
\label{Def-ComForFacH}
{\cal H}^{{\rm N}^P{\rm LO}}
=
 \!\sum_{p=0}^P
 \left(\!\frac{\alpha_s(\mu)}{2\pi}\!\right)^p
\frac{1}{2 i}\!\int_{c-i\infty }^{c+i\infty }\!\!\!dj\, \xi^{-j-1}
\frac{2^{j+1}\Gamma(5/2+j)}{\Gamma(3/2)\Gamma(3+j)} \left[\!
i+\tan\!\left(\!\frac{\pi j}{2}\!\right)\!\right]
 {\cal C}^{(p)}_j\!\!\left(\!\frac{{\cal
Q}}{\mu}\!\right)
H_j^{\rm NS}(\xi,\Delta^2,\mu^2 )
\end{eqnarray}
at NLO $(P=1)$ and NNLO $(P=2)$ accuracy. Certainly, the size of
the radiative corrections depend on the GPD distribution itself.
As a GPD model assumption, we suppose that the expansion of
$H_j^{\rm NS}$ in powers of $\xi^2$  induces a systematic
expansion of the CFF (\ref{Def-ComForFacH}). A closer look to this
issue has been given in Sect.\ 4 of Ref.\ \cite{MueSch05}. The GPD
moments are generically parameterized as
\begin{eqnarray}
\label{Def-BuiBlo} H^{\rm NS}_j(\Delta^2,\xi,\mu_0^2)= F^{\rm NS}(\Delta^2)
\frac{B(1-\alpha(\Delta^2)+j,\beta+1)}{B(1-\alpha(\Delta^2),\beta+1)}
+ {\cal O}(\xi^2)\,.
\end{eqnarray}
In the forward limit, i.e., $\Delta\to 0$,  these moments reduce
to the Mellin moments of the unpolarized parton density. Hence,
the parameters $\alpha$ and $\beta$ characterize the small and
large $x$ behavior, respectively, i.e., $q(x,\mu_0)\propto
x^{-\alpha} (1-x)^\beta$. For non-singlet parton densities the
generic values are $\alpha(\Delta^2=0)=1/2$ and $\beta=3$.  In the
off--forward kinematics we consider $\alpha(\Delta^2) = \alpha(0) +
\alpha^\prime(0)\Delta^2 $ as a linear meson Regge trajectory with
$\alpha(0) = 1/2$ and $\alpha^\prime(0) = 0.9\, \mbox{\rm
GeV}^{-2}$. Note also that the conformal GPD moment with $j=0$
reduces to the elastic form factor $F^{\rm NS}(\Delta^2)$.
Certainly, a more realistic ansatz of GPD moments is formed by a
linear combination of the building blocks (\ref{Def-BuiBlo}).

\begin{figure}[t]
\begin{center}
\mbox{
\begin{picture}(450,240)(0,0)
\put(175,240){(a)} \put(10,130){\insertfig{7.1}{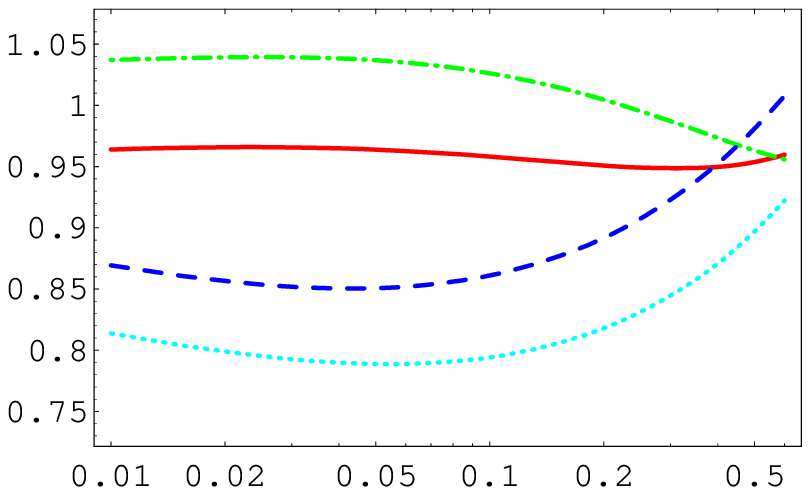}}
\put(-10,145){\rotatebox{90}{$K_{\rm abs}({\cal Q}^2= 2.5\,
\GeV^2)$}}
 \put(430,240){(b)}
\put(260,130){\insertfig{7.1}{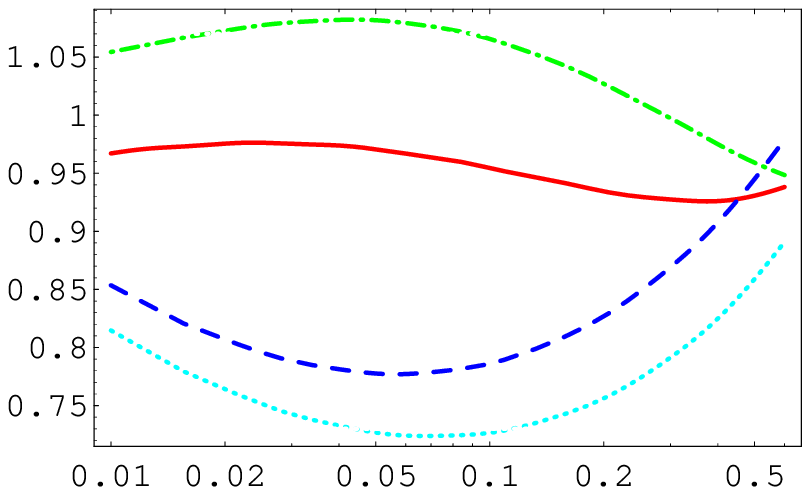}}
 \put(175,110){(c)}
\put(10,0){\insertfig{7.1}{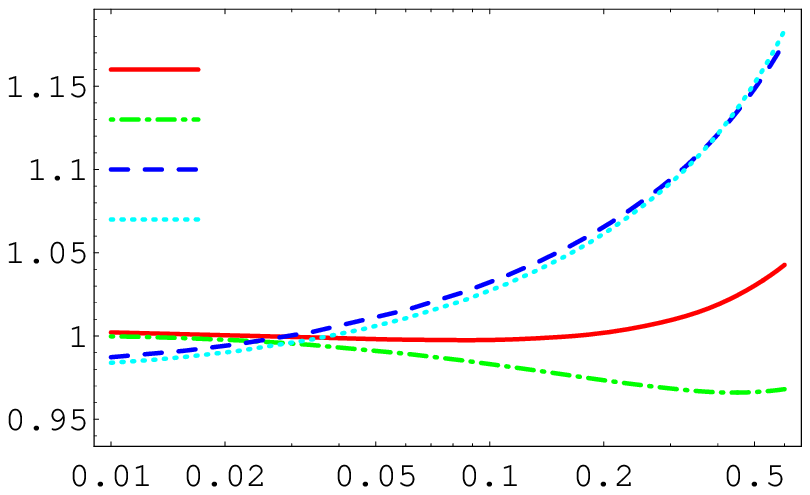}} \put(205,-5){$\xi$}
\put(430,110){(d)} \put(260,0){\insertfig{7.1}{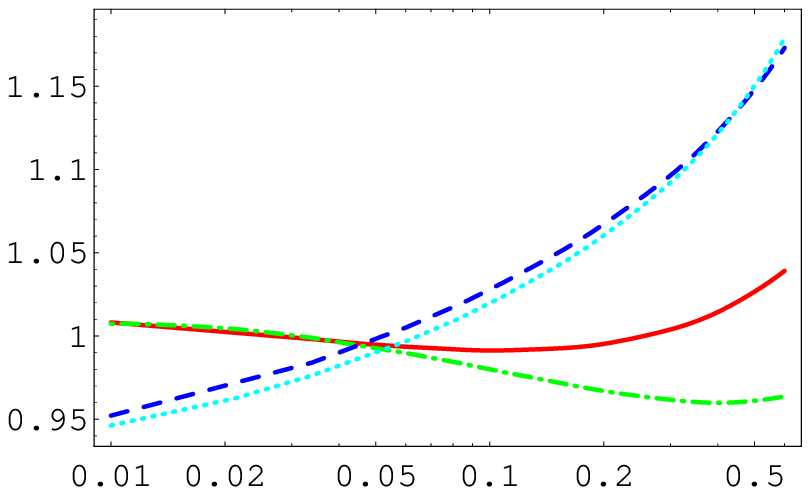}}
\put(-10,15){\rotatebox{90}{$K_{\rm arg}({\cal Q}^2= 2.5\,
\GeV^2)$}} \put(455,-5){$\xi$} \put(66,107){\tiny NNLO,\,
$\beta_0=-9$} \put(66,94){\tiny NNLO,\, $\beta_0=0$}
\put(66,82){\tiny NLO, $\overline{\rm CS}$} \put(66,69){\tiny NLO, $\overline{\rm
MS}$}
\end{picture}
}
\end{center}
\caption{ \label{FigNNLO} The relative radiative corrections,
defined in Eq.\ (\ref{Def-Rrat}), are plotted versus $\xi$ for the
modulus [(a) and (b)] and phase [(c) and (d)] of the  CFF
(\ref{Def-ComForFacH}) with $\alpha=0.5$ [(a) and (c)] and
$\alpha= -0.1$ [(b) and (d)]: NNLO in full (solid) and $\beta_0=0$
(dash--dotted) as well as in NLO for the $\overline{\rm CS}$ (dashed) and
$\overline{\rm MS}$ (dotted) scheme. We  always set $\mu=
{\cal Q}$ and  $\alpha_s({\cal Q}^2= 2.5\, \GeV^2)/\pi = 0.1$. }
\end{figure}
Let us first compare the radiative corrections in the
$\overline{\rm MS}$ and $\overline{\rm CS}$ scheme to NLO accuracy. Strictly
spoken, at a given input scale ${\cal Q}_0$ there is no difference
between both predictions, if the non--perturbative quantities are
transformed, too,
and a consequent expansion in $\alpha_s$ is performed.  Usually,
the GPD moments are taken from some non--perturbative (model)
calculation or ansatz, where the matching with the perturbative
prescriptions has its own uncertainties. So let us take the same
input in both schemes and study the relative changes
\begin{eqnarray}
\label{Def-Rrat} K^P_{\rm abs}=\frac{\left|{\cal H}^{{\rm N}^P{\rm
LO}}\right|}{\left|{\cal H}^{{\rm N}^{P-1}{\rm LO}}\right|}\,,
\qquad {K}^P_{\rm arg}= \frac{{\rm arg}\!\left(\! {\cal H}^{{\rm
N}^P{\rm LO}}\!\right)}{{\rm arg}\!\left( {\cal H}^{{\rm
N}^{P-1}{\rm LO}}\right)}
\end{eqnarray}
to the modulus and phase of the CFF (\ref{Def-ComForFacH}). These
quantities do not suffer from large radiative corrections as it is
artificially the case for the real part of the
amplitude\footnote{\label{Foo-DefRadCor}The real part possesses a
zero in the valence quark region, which position depends on  the
approximation. Hence, in the vicinity of this zero the relative
radiative corrections blow up.}. One should bear in mind that
these factors are a measure for the necessary reparameterization
of the GPD when one includes the next order in a given scheme. In
Fig.\ \ref{FigNNLO} we depict for the typical kinematics in fixed
target experiments, i.e., $0.05\lesssim \xi \lesssim 0.3 $, the
$K$ factors to NLO as dashed and dotted lines for the
$\overline{\rm CS}$ and $\overline{\rm MS}$ schemes, respectively.
We set $\mu={\cal Q }$ and independent of the considered
approximation we choose $\alpha_s(\mu_r^2= 2.5\,{\rm GeV}^2 ) =
0.1 \pi$. From the panels (a) and (b) it can been realized that in
the $\overline{\rm MS}$ scheme the radiative corrections to the
modulus are up to 20\% and 30\% for $\alpha=0.5$ ($\Delta^2=0$)
and $\alpha=-0.1$ ($\Delta^2=-0.\overline{6}\; {\rm GeV}^2 $),
respectively. In the $\overline{\rm CS}$ scheme these radiative
corrections  are reduced by 30\%. Note that such a reduction has
been observed in a quite different processes, namely, for the
photon--to--pion transition form factor\footnote{This process is
also evaluated within the OPE and the  reduction of NLO
corrections has a common origin. Namely, in the $\overline{\rm
CS}$, compared to the $\overline{\rm MS}$, scheme the first few
Wilson--coefficients are smaller, while the stronger logarithmic
growing with increasing $j$ is anyway suppressed by the
non--perturbative input, see Eqs.\ (\ref{Res-WilCoe-CS-NLO}) and
(\ref{Res-WilCoe-MS-NLO}).}. The relative radiative corrections to
the phase is in both cases of about 15\% at $\xi=0.6$ and
diminishes with decreasing $\xi$. These findings qualitatively
agree with previous ones in which the Radyushkin ansatz for GPDs
was used \cite{BelMueNieSch99}.

We study now the radiative corrections to NNLO accuracy. To
simplify their evaluation, we take for $c_j^{(2)}$ a fit, given in
Ref.\ \cite{NeeVog99}, rather than the exact expression. For the
$\beta_0$ proportional term we have checked that within
$|c|=|\Re{\rm e} j| \le 1/2$, see Mellin--Barnes integral
(\ref{Def-ComForFacH}), the accuracy is on the level of one per
mill or better. Outside this region, the deviation can be larger
and one might use Fortran routines \cite{BluMoc05}.  For three
quark flavors $n_f=3$, the same scale settings, and the initial
value of $\alpha_s$ as specified above for NLO,  we plot in Fig.\
\ref{FigNNLO} the ratios (\ref{Def-Rrat}) in the $\overline{\rm
CS}$ scheme for $P=2$ as solid line. The $\xi$--dependence of the
modulus $K$--factor, see panels (a) and (b), is rather flat and
the modulus decreases on the 5\% level. The radiative correction
to the phase is again negligible for smaller values of $\xi$ and
increases now only to 5\% at $\xi=0.6$. We note that the $\beta_0$
induced corrections are about twice times larger than the
remaining ones and are opposite in sign, see dash--dotted line.
Remarkably, for $\beta_0=0$ we find then  an opposite behavior of
the $K$ factors as in NLO. This arise from a sign alternating
series and as for the photon--to--pion transition form factor, this
might be  considered as a reminiscence on Sudakov double logs
\cite{MusRad97}.

Let us finally address the modification of the scale dependence
due to the higher order corrections. Note that the analysis about scheme
dependence in Ref.\ \cite{Mue98} suggests that the discrepancy
between the $\overline{{\rm CS}}$ and $\overline{{\rm MS}}$
schemes at NLO accuracy are mainly induced by the
Wilson--coefficients while the evolution yields minor differences.
So we only consider here the $\overline{{\rm CS}}$ scheme and
analogously  as in  Eq.\ (\ref{Def-Rrat}), we quantify the
relative changes of $d{\cal H}/d\ln{\cal Q}^2$ by the ratios
\begin{eqnarray}\label{Def-Rrat-dot}
\dot{K}^P_{\rm abs}= \left|\frac{d {\cal H}^{\rm {N}^P{\rm
LO}}}{d\ln{\cal Q}^2}\right|\Bigg/\left|\frac{d {\cal H}^{\rm
{N}^{P-1}{\rm LO}}}{d\ln{\cal Q}^2}\right|\,, \; \dot{K}^P_{\rm
arg}=
  \left[\pi +{\rm arg}\!\left(\!
  \frac{d\, {\cal H}^{\rm {N}^P{\rm LO}}}{d\ln{\cal Q}^2}\!\right)
  \right]\Bigg/
  \left[\pi +{\rm arg}\!
  \left(\! \frac{d\, {\cal H}^{\rm {N}^{P-1}{\rm LO}}}{d\ln{\cal
  Q}^2}\!\right)\right].
\end{eqnarray}
Both the numerator and denominator in the latter $\dot{K}$--factor
are now defined in the interval $[0,2\pi]$ and so the appearance
of a zero in the denominator is avoided, cf.\ footnote
\ref{Foo-DefRadCor}. We take the same scale setting and initial
condition for the (exact) evolution of $\alpha_s({\cal Q})$ as
above. The conformal moments (\ref{Def-BuiBlo}) are evolved in the
$\overline{\rm CS}$ scheme.  The input scale is now $\mu_0^2= 0.5\,
{\rm GeV}^2$, which is typical for the matching of perturbative
and non--perturbative QCD. The non--leading logs in the solution of
the evolution equation (\ref{Def-RGE}) are expanded with respect
to $\alpha_s$ and are consistently combined with the
Wilson--coefficients (\ref{Res-WilCoe-Exp-CS}) in the considered
order. Here the forward anomalous dimensions are taken from Ref.\
\cite{MocVerVog04} and the unknown NNLO mixing term
$\Delta_{jk}^{\overline{{\rm CS}}}$ in Eq.\ (\ref{Def-RGE}) is
neglected. This mixing can be suppressed at the input scale within
an appropriate initial condition  and so we expect only a minor
numerical effect; see also Ref.\
Ref.\ \cite{Mue98}.
\begin{figure}[t]
\begin{center}
\mbox{
\begin{picture}(450,120)(0,0)
 \put(175,100){(a)}
\put(10,0){\insertfig{7.2}{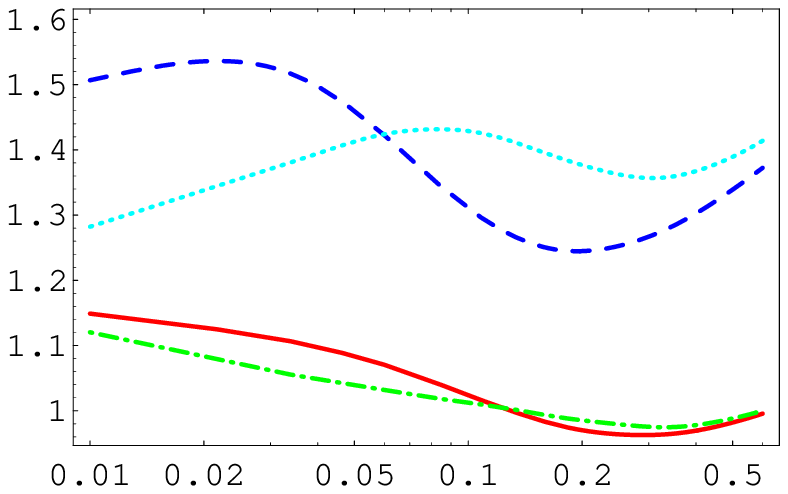}} \put(205,-5){$\xi$}
\put(425,100){(b)} \put(260,0){\insertfig{7.1}{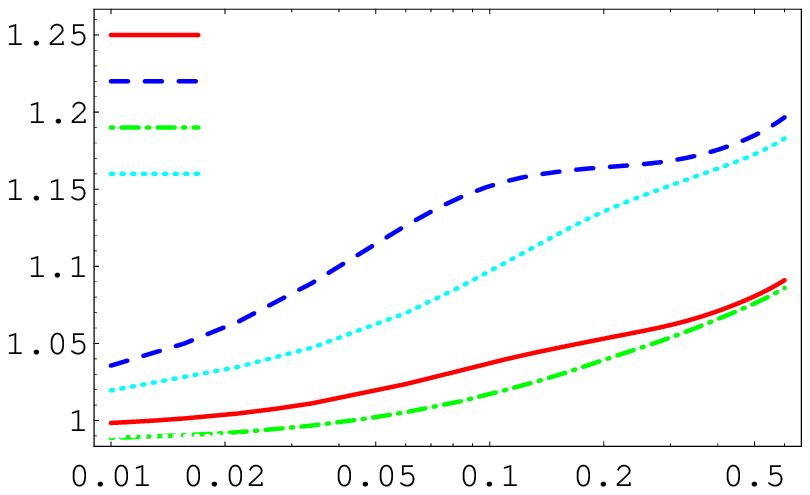}}
\put(-15,20){\rotatebox{90}{$\dot{K}^{\rm abs}({\cal Q}^2= 4\,
\GeV^2 ) $}} \put(235,20){\rotatebox{90}{$\dot{K}^{\rm arg}{\cal
Q}^2=4\,\GeV^2 ) $}} \put(455,-5){$\xi$} \put(315,116){\tiny
NNLO,\, $\alpha=0.5$} \put(315,104){\tiny NLO,\,\phantom{N}
$\alpha=0.5$} \put(315,92){\tiny NNLO,\,
 $\alpha=-0.1$} \put(315,81){\tiny NLO,\,\phantom{N}
$\alpha=-0.1$}
\end{picture}
}
\end{center}
\caption{ \label{Fig-ScaDep} The relative change of scale
dependence, cf.\ Eq.\ (\ref{Def-Rrat-dot}), in the $\overline{\rm
CS}$ scheme at NLO (dashed, dotted)  and NNLO (solid, dash--dotted)
versus $\xi$ is depicted for the modulus (a) and phase (b) of the
CFF (\ref{Def-ComForFacH}) with $\alpha=0.5$ (dashed, solid)  and
$\alpha= -0.1$ (dotted, dash--dotted) and ${\cal Q}^2 =4\, \GeV^2$.
We set $\mu = {\cal Q}$, $\alpha_s(\mu_r^2= 2.5\, \GeV^2 ) /\pi =
0.1$ and took the input (\ref{Def-BuiBlo}) at the scale ${\cal
Q}^2_{0} = 0.5\, \GeV^2 $. }
\end{figure}
The dashed and dotted lines in Fig.\ \ref{Fig-ScaDep} show that in
NLO the scale dependence changes of about 30\% to 50\% for the
modulus and up to 20\% for the phase. The latter $\dot{K}$--factor
becomes close to one for smaller values of $\xi$. The NNLO
corrections, compared to the full NLO result, are milder. We
observe (solid and dash--dotted line) a 20\% or less and a maximal
10\% correction for the modulus and phase, respectively.

\section{Conclusions}
\label{Sec-Con}

In this letter we have taken the first steps in the full investigation
of radiative NNLO corrections to the DVCS scattering
amplitude. Thereby, we employed the COPE, which allows an
economical treatment of perturbative corrections, and restricted
ourselves to the CFF $\cal H$ in the flavor non--singlet sector.
This study can be immediately adopted for $\cal E$. The extension
to the axial vector case, i.e., to $\widetilde {\cal H}$ and
$\widetilde {\cal E}$, is straightforward, however, here the anomalous
dimensions to three loop order cannot be borrowed from the
polarized DIS results. The conformal  approach can be also
straightforwardly extended to the flavor singlet sector, which
will be presented somewhere else.

We relied on GPDs for which the expansion of the conformal moments
in powers of $\xi^2$ is meaningful and used in our analysis only
the leading term. Whether this assumption is  too restrictive is
an open problem, which might be resolved by lattice calculations
\cite{Hagetal03}. At least there is a warning from perturbative
QCD. Namely, in the $\overline{\rm MS}$ scheme $\xi^2$--suppressed
contributions in the Wilson--coefficients are resummed and lead to
larger radiative corrections than in the $\overline{\rm CS}$ one.
Let us summarize our numerical findings for fixed target
kinematics, i.e., $0.05 \lesssim\xi \lesssim 0.3$:
\begin{itemize}
\item Compared to
the $\overline{\rm MS}$ scheme, in the $\overline{\rm CS}$ one the
radiative NLO corrections to the modulus of the CFF are reduced by
30\% or so, while for the phase no significant differences appear.

\item The relative NNLO corrections compared to the NLO ones are
of the 5\% level or below, where the $\beta_0$ proportional
ones dominate and determine the sign.

\item The change of the  scale dependence due to the NLO corrections
is rather large and is for the modulus (phase) of the CFF of about
30\% to 40\% (15\% or smaller). Comparing NNLO to NLO, the
two--loop corrections are reduced to less than 20\% (10\%) .
\end{itemize}
Let us add that the
Wilson--coefficients in the vector and axial--vector channel  possess a
similar $j$ dependence. So we might expect that for
$\widetilde {\cal H}$  and $\widetilde {\cal E}$ (axial--vector
case) comparable radiative corrections appear.
We conclude that our findings support the perturbative description
of the DVCS process at scales which are accessible in the
kinematics of present fixed target experiments.

{\ }

\noindent
{\bf Acknowledgement}

\noindent
For discussions I am indebted to A.\ Manashov, B.~Meli{\' c}, K.~Kumeri{\v c}ki,
K.~Passek--Kumeri{\v c}ki,  and A.\ Sch\"afer. This project has been
supported by the Deutsche Forschungsgemeinschaft and the
U.S. National Science Foundation under grant no. PHY--0456520.

\end{document}